\pdfoutput=1
\documentclass[twocolumn]{aastex63}
\usepackage{apjfonts}
\usepackage{graphicx}
\usepackage{xspace}
\usepackage{enumitem}
\usepackage{amsmath}

\begin{document}

\lefthead{{\em \textit{\textit{NuSTAR}}, \textit{Swift/XRT}} Observations of MAXI J0637$-$430}
\righthead{Lazar et al.}

\def\lsim{\mathrel{\lower .85ex\hbox{\rlap{$\sim$}\raise
.95ex\hbox{$<$} }}}
\def\gsim{\mathrel{\lower .80ex\hbox{\rlap{$\sim$}\raise
.90ex\hbox{$>$} }}}

\title{Spectral and Timing Analysis of {\em \textit{NuSTAR}} and {\em \textit{\textit{Swift/XRT}}} Observations of the X-Ray Transient MAXI J0637$-$430}

\author{Hadar Lazar}
\affiliation{Space Sciences Laboratory, 7 Gauss Way, University of California, Berkeley, CA 94720-7450, USA}

\author{John A. Tomsick}
\affiliation{Space Sciences Laboratory, 7 Gauss Way, University of California, Berkeley, CA 94720-7450, USA}

\author{Sean N. Pike}
\affiliation{Cahill Center for Astronomy and Astrophysics, California Institute of Technology, Pasadena, CA 91125, USA}

\author{Matteo Bachetti}
\affiliation{INAF - Osservatorio Astronomico di Cagliari, via della Scienza 5, I-09047 Selargius, Italy}

\author{Douglas J.K. Buisson}
\affiliation{Department of Physics and Astronomy, University of Southampton, Highfield, Southampton, SO17 1BJ, UK}

\author{Riley M. T. Connors}
\affiliation{Cahill Center for Astronomy and Astrophysics, California Institute of Technology, Pasadena, CA 91125, USA}

\author{Andrew C. Fabian}
\affiliation{Institute of Astronomy, University of Cambridge, Madingley Road, Cambridge CB3 0HA, UK}

\author{Felix Fuerst}
\affiliation{European Space Astronomy Centre, E-28691 Villanueva de la Cañada, Madrid, Spain}

\author{Javier A. Garc\'ia}
\affiliation{Cahill Center for Astronomy and Astrophysics, California Institute of Technology, Pasadena, CA 91125, USA}
\affiliation{Dr. Karl Remeis-Observatory and Erlangen Centre for Astroparticle Physics, Sternwartstr.~7, 96049 Bamberg, Germany}

\author{Jeremy Hare}
\affiliation{NASA Goddard Space Flight Center, Greenbelt, MD 20771, USA}
\affiliation{NASA Postdoctoral Program Fellow}

\author{Jiachen Jiang}
\affiliation{Department of Astronomy, Tsinghua University, Shuangquing Road 30, Beijing 100084, China}

\author{Aarran W. Shaw}
\affiliation{Department of Physics, University of Nevada, Reno, NV 89557, USA}

\author{Dominic J. Walton}
\affiliation{Institute of Astronomy, University of Cambridge, Madingley Road, Cambridge CB3 0HA, UK}

\begin{abstract}

We present results for the first observed outburst from the transient X-ray binary source MAXI J0637$-$430. 
This study is based on eight observations from the \textit{Nuclear Spectroscopic Telescope Array} (\textit{NuSTAR}) and six observations from the \textit{Neil Gehrels Swift Observatory X-Ray Telescope} (Swift/XRT) collected from 2019 November 19 to 2020 April 26 as the 3--79 keV source flux declined from $8.2 \times 10^{-10}$ to $1.4 \times 10^{-12}$  \,erg\,cm$^{-2}$\,s$^{-1}$. We see the source transition from a soft state with a strong disk-blackbody component to a hard state dominated by a power-law or thermal Comptonization component. \textit{NuSTAR} provides the first reported coverage of MAXI J0637$-$430 above 10 keV, and these broadband spectra show that a two-component model does not provide an adequate description of the soft state spectrum. As such, we test whether blackbody emission from the plunging region could explain the excess emission. 
As an alternative, we test a reflection model that includes a physical Comptonization continuum. Finally, we also test a spectral component based on reflection of a blackbody illumination spectrum, which can be interpreted as a simple approximation to the reflection produced by returning disk radiation due to the bending of light by the strong gravity of the black hole. We discuss the physical implications of each scenario and demonstrate the value of constraining the source distance.

\end{abstract}


\section{Introduction}

A low-mass X-ray binary (LMXB) consists of a low-mass donor star accreting matter onto a compact object that is either a black hole or neutron star. As there are only under two hundred catalogued LMXBs \citep{catalogue} in the Galaxy and the Magellanic Clouds, they provide a unique window into the endpoint of stellar evolution. Of particular interest is the interplay between the strong gravity of these elusive compact objects and their respective time-dependent accretion disks. Phase-resolved spectroscopy can be used to understand the processes that occur close to the neutron star surface or the black hole horizon.  

The outer region of an accretion disk around the compact object typically cycles between a neutral, quiescent state and an ionized, outburst state. According to the disk instability model (DIM) \citep{dim,dim2}, the accumulating matter on the thin accretion disk heats and ionizes a significant portion of the disk and causes a thermal-viscous instability. The increased viscosity of the disk in an ionized state causes a greater outward redistribution of angular momentum, which is thought to be physically mechanised by the magneto-rotational instability \citep{balbus, tet2}. This sharply increases the accretion rate, which in turn causes the X-ray luminosity to rise several orders of magnitude above the quiescent state. As the timescale of this bright outburst lasts from days to months, it provides an opportunity for X-ray telescopes to study the poorly understood accretion process.

A typical LMXB has notable spectral features.
The accretion disk surrounding the compact object has opacities that are dominated by free-free absorption, and can be modeled as a disk blackbody \citep{shak}. The soft X-ray photons originating from the standard accretion disk undergo inverse-Comptonization on hot electrons in the corona, and produce the hard power-law component \citep{haa, zdz2}. Thus the two spectral states can be approximated with (1) a soft thermal multi-colored blackbody component and (2) a hard non-thermal power-law component. Additionally, they show intermediate behavior as the sources transition between these
states.  

Although both types of compact objects may include these components, in neutron stars, the presence of a hard surface means that excess energy of accretion is released upon impact and radiated to infinity. This manifests in the neutron star surface having a high-temperature blackbody component. It is for this reason that black holes are associated with ultrasoft accretion with a lower-temperature thermal component. Generally, the two components that inform this are detectable in soft X-rays ($<$ 10 keV).

The previous analyses in this energy band of the 2019-2020 outburst of Black Hole Candidate (BHC) MAXI J0637$-$430 \citep{tet, jana} demonstrate that the spectral properties are consistent with those of a black hole X-ray binary (BHXRB). \citet{jana} report no signs of high-frequency kHz quasi-periodic oscillations, which are seen in
several NS LMXBs. Power density spectra (PDS) are also provided and show that the power decreased rapidly at frequencies $> 10$ Hz, in accordance with other BHXRBs. They also attempt to add an emission component for potential emission from the surface of a neutron star, but found it to be insignificant in their energy band. Though these results favor a black hole as the compact object, they do not rule out a neutron star.
 
We analyze the broadband spectra of MAXI J0637$-$430 covering up to 79 keV and report excess emission beyond the two-component model revealed in the soft state. This additional component may be attributed to blackbody emission from the plunging region, i.e. the perceived "boundary" of the black hole itself, which has been identified in a recent study of MAXI J1820+070 \citep{plunge}. This could be modeled with an additional blackbody component associated  with  the  innermost  part  of  the  disk and  with  the  start  of  the  plunging  region.  The reasoning for this is that, while historically models have assumed a zero-stress boundary condition at the Innermost Stable Circular Orbit (ISCO) \citep{nov, rem}, later work suggested that magnetic stresses may occur. These stresses at the ISCO may temporarily delay the plunge of infalling plasma, allowing further energy release and keeping the plasma optically thick to provide additional blackbody flux beyond what is included in standard disk models. 

We also consider a reflection spectrum \citep{refl} in identifying the extra emission. In many models, the reflected radiation is thought to stem from the reprocessing of
high-energy coronal photons in the optically thick accretion
disk. The resulting spectrum includes radiative recombination
continua, absorption edges, and nuclear emission lines. By modeling the
reflection spectrum, one can estimate both the accretion disk inclination 
and the black hole spin parameter by determining the broadening and shape of emission lines. 

A common model of this reflection component includes (1) calculation of the reprocessed emission that arises from illumination of the disk by the high-energy Comptonized continuum in the rest-frame of the disk and (2) the full ray tracing calculations from the
irradiating source to the disk and onward to the observer (\texttt{relxill}; \citealt{daus, Garc}). The former feature produces line emissions, notably the iron line emissions, as well as Compton down-scattering of higher energy photons. The latter includes the calculation of the relativistic effects which distort the spectrum, such as lightbending effects, Doppler shifts, and gravitational redshifts. However, some previous studies (e.g. the study of BHXRB XTE J1550$-$564; \citealt{ex}, BHXRB 4U 1630$-$47 outburst; \citealt{con}) have found that the coronal inverse-Compton (IC) reflection model does not adequately describe the reflection continuum, and it is necessary to have an alternative model which adopts
a softer, thermal continuum as its irradiating spectrum. It was found that the reflection spectrum in the very soft state is best explained by disk self-irradiation, meaning photons from the inner disk are bent by the strong gravity of the black hole and reflected off the disk surface. Despite the theoretical framework of returning radiation being spearheaded decades ago \citep{cunning}, a full relativistic model for returning thermal radiation reprocessed in the disk has not been produced. Instead, the authors have adopted an approximation to this scenario by implementing reflection models produced with an illuminating blackbody spectrum instead of the standard power-law spectrum typically assumed. This model, called relxillNS (García et al. 2021, submitted), was originally designed to treat reflection in accreting neutron stars, and thus it does not explicitly include the light-bending effects that produce returning radiation. However, it does serve as a proxy of what the reflection signature might look like under these circumstances. Presently, theoretical efforts are underway to produce a self-consistent model for returning radiation around black holes (e.g., Dauser et al. 2021, submitted).

 These components are detectable by telescopes that provide
broadband X-ray coverage, sensitivity to the soft emission, and moderate to high spectral resolution. We therefore conducted a broadband X-ray campaign with \textit{Neil Gehrels Swift Observatory X-Ray Telescope} (XRT) \citep{bur} and \textit{Nuclear Spectroscopic Array} (\textit{NuSTAR}) \citep{har} for the LMXB BHC MAXI J0637$-$430. Given \textit{NuSTAR}'s ability to simultaneously observe the iron emission line and the Compton hump without pileup effects, it is effective for studying reflection spectra in LMXBs. As \textit{NuSTAR}'s hard X-ray continuum (3--79 keV) allows us to identify the presence of Comptonization, and  XRT's soft X-ray continuum (0.5--10 keV) is effective in identifying a disk blackbody component, these instruments together allow us to identify all the components needed to characterize the LMXB.  Moreover, the components can be used to calculate the range of the inner disk radius, which is used to understand the geometry and physics of the LMXB. 

When the disk-blackbody component is no longer detected towards the end of the outburst, we turn to characteristic break frequencies obtained from the Power Density Spectrum, as they may also share a relationship with the inner radius of the accretion disk. This is because as the mass accretion rate drops, the inner radius of a disk ($R_{\rm in}$) increases, causing a drop in the dynamical timescales, and consequently, the characteristic frequencies for the system \citep{dimatteo}. We can subsequently place an upper bound on the increase of the inner radius from Keplerian orbits, which are the fastest variability timescale at any radius around a compact object. Break frequencies are then compared to changes in the power-law photon index, which previous black hole system studies in the hard state have found to be correlated \citep{gil, rev, kal, freq}.

 The following section outlines the data reduction of the LMXB MAXI J0637$-$430. Section 3 tests whether the spectra could be adequately described by a model with emission from the plunging region or a model with a reflection component. Section 4 details analysing the Power Density Spectrum. Lastly, we discuss our results and their physical implications.

\section{Observation and Data Reduction}

\begin{figure*}[ht]
\centering
\includegraphics[width=0.7\textwidth]{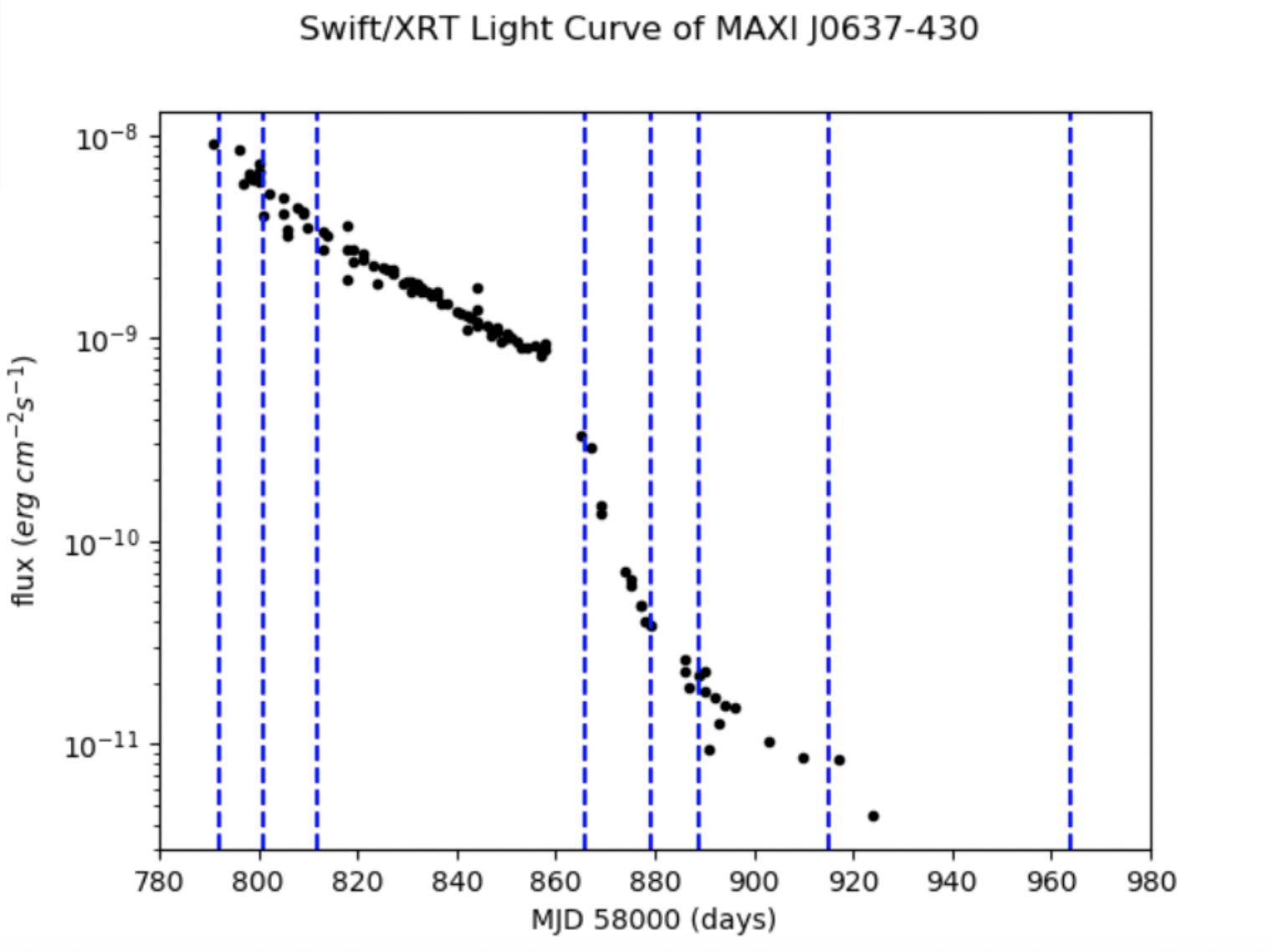}
\caption{\textit{Swift/XRT} fluxes for MAXI J0637$-$430, from \citet{tet}. The dashed lines indicate the dates \textit{NuSTAR} observations were made.}
\label{fig:light} 
\end{figure*}

MAXI J0637$-$430 was discovered on 2019 November 2 (MJD 58789) \citep{gsc} by MAXI/GSC. Following the discovery, \textit{NuSTAR} made eight observations from November 5, 2019 to April 26, 2020, six of which were observed contemporaneously with XRT. This six-month multi-wavelength campaign ended when XRT made a non-detection in the X-ray band on 2020 June 11 \citep{nondet}. In Table 1, we detail the exposure times and source flux evolution of this multi-wavelength campaign through the soft, intermediate, and hard states. In Figure \ref{fig:light}, we provide the XRT light curve for the outburst.

\subsection{\textit{NuSTAR}}

All \textit{NuSTAR} data were processed using calibration data base (CALDB) files from 2019 September 10 and analyzed using XSPEC version 12.11.1c \citep{xspec}. We used the \texttt{nupipeline} tool (version v0.4.6)
to filter the event lists, and then the \texttt{nuproducts} tool to extract
spectra and response files for the focal plane modules A and B (FPMA and FPMB). For the first six observations, the source and background extraction region was a circle of radius 86". As by the last two observations the outburst dimmed, the radii were reduced to 32" for 80502324014 and 27" for 80502324016. The FPMA and FPMB spectral data were generated as two different groups using GRPPHA in HEASOFT version 6.27.2 (HEASARC 2014) with a signal-to-noise ratio of 10 in each bin for the first five observations, a signal-to-noise ratio of 5 for 80502324012 and 80502324014, and a signal-to-noise ratio of 4 for 80502324016. A XRT observation was added as a third group when the observations coincided. These groups were then jointly fit using models available in XSPEC and the relxill suite, which provides self-consistent
relativistic disk reflection models \citep{daus, Garc}. We also used \texttt{relxillNS}, a recent development of the relxill suite, which gives the reflection of the disk illuminated by the neutron star surface or black hole boundary layer. It assumes a blackbody is irradiating the disk, rather than the power-law given by the traditional \texttt{relxill} model.

\subsection{\textit{Swift/XRT}}
We extracted 0.5--10\,keV energy spectra for the six \textit{Swift/XRT} observations that occurred during \textit{NuSTAR} observations (Table 1).  These observations have relatively short exposure times, covering only a fraction of the much longer \textit{NuSTAR} observations.  We extracted the spectra using Swift data analysis tools in HEASOFT v6.27.2 with the use of CALDB files.  The spectra were grouped using a signal-to-noise ratio of 5 in each bin for all observations except the last one of 00012172093, which had a signal-to-noise ratio of 2. For all of the reductions described below, we only used events flagged with grade 0.

The XRT count rate changed significantly from 175\,c/s in the first observation to 0.05\,c/s in the last observation.  Thus, different XRT instrument modes were used as well as different extraction methods for the spectra.  For the first four observations, XRT was in windowed timing (WT) mode, and for the first two observations (ObsIDs 00012172008 and 00012172018), the count rate was high enough for photon pile-up to be a concern.  Thus, for these two observations, the source extraction region was an annulus with inner radius of $20^{\prime\prime}$ and outer radius of $47^{\prime\prime}$.  A background spectrum was also extracted from an active detector region far from the source, and the background level was scaled based on the relative sizes of the source and background regions.  We used the appropriate response file (swxwt0s6\_20131212v015.rmf) and used {\ttfamily XRTmkarf} to produce the ancillary response file.  For the other two WT observations (ObsIDs 00012172066 and 00012172077), the count rates are lower, and the source extraction region was a circle of radius $47^{\prime\prime}$ rather than an annulus.

XRT was in photon counting (PC) mode for the final two \textit{Swift/XRT} observations and the count rates were 0.28\,c/s and 0.05\,c/s.  At these low count rates, photon pile-up is not a concern, and the extraction region we used was a circle of radius $47^{\prime\prime}$.  For these spectra, the appropriate response file is swxpc0s6\_20130101v014.rmf, which is the file we used.  For all six observations, we used an exposure map when making the ancillary response file.  It is important to do this in cases where there are bad pixels or pixel rows in the source region.  We especially point this out here because the low count rate for the final observation (ObsID 00012172093) was partly due to the fact that the source was on a row of bad pixels.
\newline

\section{Spectral Results}

We carried out spectral studies of the BHC MAXI J0637$-$430 during
its 2019-2020 X-ray outburst using observations from \textit{NuSTAR} in the 3--79 keV range and observations from XRT in the 0.5--10 keV range. For the joint fitting between XRT, FPMA, and FPMB, a cross-normalization constant is allowed
to vary freely with the convention that FPMA is unity. The value of the cross-normalisation factor between FPMA and FPMB is found to always lie within \textit{NuSTAR}'s accepted limit of $\leq$ 5 percent \citep{mad}. The neutral hydrogen column density ($N_{\rm H}$) is fixed to $4.39 \times 10^{20}$ cm$^{-2}$ \citep{tet}, which was given by the reddening E(B-V) $\sim$ 0.064 in
\citet{nh}, for all the described models. It is noteworthy that relaxing this constrain on the column density appeases low-energy residuals (see Appendix for more details). All parameter uncertainties are reported at the $90$ percent confidence level for one
parameter of interest. Furthermore, all
models described include the Galactic absorption through the implementation of the \texttt{TBabs} model \citep{tbabs}.

The following two subsections report on tests conducted to determine which model best describes the soft-state spectra. We evaluated the quality of the different models by comparing their $\chi^{2}$ values when fit to the robust  observations made on MJD 58801 and 58812. The results obtained from observations made on MJD 58812 are provided below. 
\newline
\newline
\subsection{Soft-state spectral results without reflection}

We first fit the soft-state
 spectra (see Table 1 for state characterization of each observation) with an absorbed thermal Comptonization model \citep{zdz}, \texttt{TBabs x Nthcomp} in XSPEC notations. The \texttt{Nthcomp} component describes the continuum shape from the thermal comptonization of photons by a cloud of electrons (the "corona"). The presence of positive residuals in soft X-ray ranges, specifically in
the rising phase of the outburst,
motivated us to add a multi-color blackbody component \citep{mit, mak} to account for the direct photons from the accretion disk. This two-component model, comprising of a combination of multi-color disk black-body
and thermal Comptonization component, \texttt{TBabs x (diskbb + Nthcomp)}, provides a better explanation of the observed broadband energy spectra.  As the \texttt{Nthcomp} component accounts for the low-energy rollover due to the seed photons from the accretion disk, the \texttt{Nthcomp} seed photon temperature and the \texttt{diskbb} inner disk temperature were tied together (kT$_{\rm in}$).  We note that for the three observations collected for the soft state, the one occurring on MJD 58792 did not have a simultaneous XRT observation. We therefore omit the \texttt{diskbb} parameter values for this observation, as without the features in the XRT energy range this component is not well-constrained.

\begin{figure}[ht]
\centering
\includegraphics[width=0.50\textwidth]{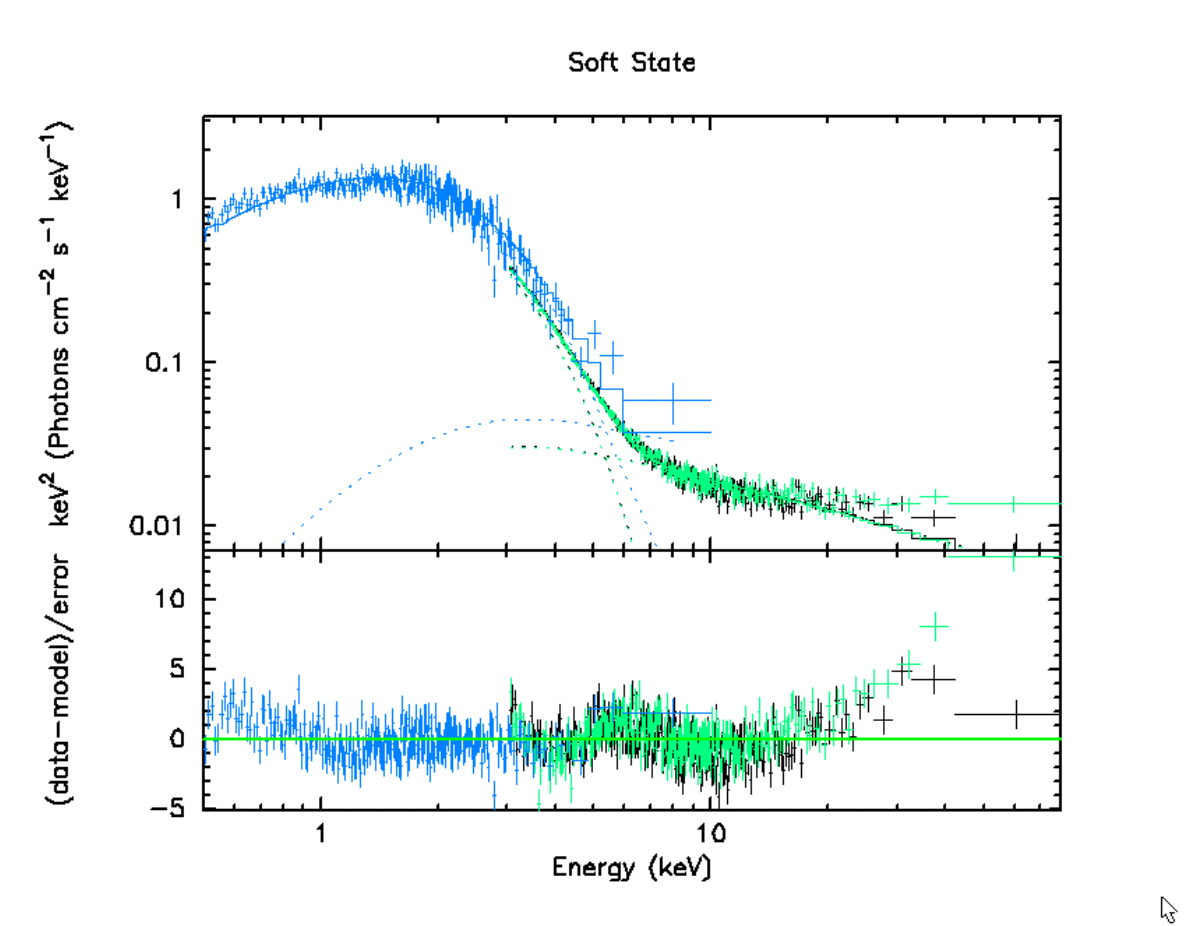}
\caption{Spectra for \textit{NuSTAR} ’s 80502324006 observation (black for FPMA, green for FPMB) and XRT’s 00012172018 observation (blue) fitted with a disk blackbody and a thermal Comptonization component. The fit produces positive residuals at the highest energies, near 6 keV, and below 1 keV.}
\label{fig:006_i}
\end{figure}

It is notable that an X-ray campaign recording up to the 10 keV X-ray range reported that the XRT data are well described by two-component model of a disk-blackbody and absorbed power law \citep{tet}. However, we find that once incorporating the higher energy range of \textit{NuSTAR}, a two-component model does not sufficiently describe the soft-state observations, giving a $\chi^{2}$ of 1242 for 720 degrees of freedom (dof) for \texttt{TBabs x (diskbb + Nthcomp)}. Particularly, we see positive residuals near 6 keV, below 1 keV, and notably at the higher energies (see Figure \ref{fig:006_i}). We therefore tested whether (i) blackbody emission from the plunging region (ii) reflection of the high energy component, or (iii) reflection of a thermal returning radiation component, or a combination of these provide the best explanation of these spectra. 

\begin{figure*}[ht]
\centering
\includegraphics[width=1\textwidth]{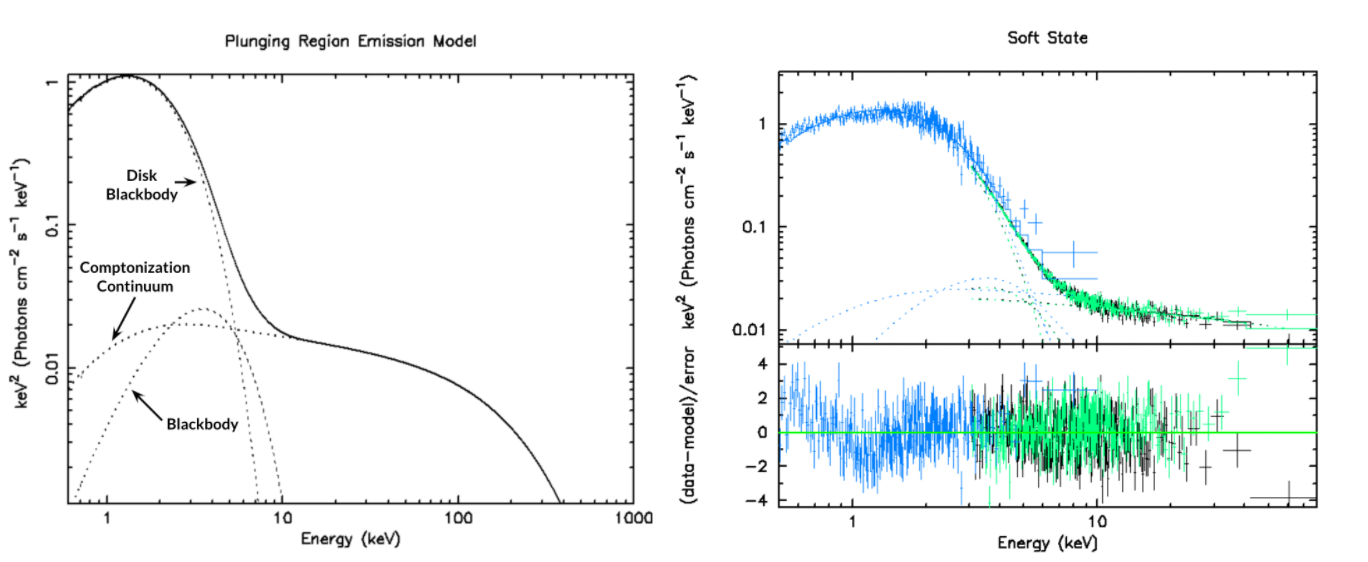}
\caption{\label{fig:plunge} (a) The model consisting of components for a disk blackbody, thermal Comptonization, and emission from the plunging region. (b) Spectra for \textit{NuSTAR} ’s 80502324006 observation and XRT’s 00012172018 observation fitted with components for a disk blackbody, thermal Comptonization, and additional blackbody component.}
\end{figure*}

First, we tested the addition of a single-temperature blackbody component, \texttt{TBabs x (diskbb + Nthcomp + bbodyrad)}. This introduces an extra blackbody component associated with the boundary of the black hole, interpreted as the start of the plunging region. This is motivated by the fact that recent work suggests \citep{plunge} that plasma may remain optically-thick slightly within the ISCO, powering the additional emission. When a model accounting for this emission is fitted, we obtain a $\chi^{2}$ of 859 for 718 dof (see Fig. \ref{fig:plunge}).

To check for physical consistency of the plunging region scenario, we can use the normalization of the blackbody in order to estimate the width of the ring at the ISCO and determine if it is a physically plausible value. In order to proceed, we make the reasonable assumption that the disk extends to the ISCO, giving us $R_{\rm ISCO}$ between $1-6 \; R_{\rm g}$, in which $R_{\rm g}$ is the gravitational radius. This is consistent with the fact that the plunging region model is thought to be most relevant for low-spin black holes, which have $R_{\rm ISCO}$ close to $6 R_{g}$. Using $R_{\rm g}=GM/c^{2}$ and taking a BH mass of 10 $M_\odot$, we find that the $R_{\rm ISCO}$ falls between 15 to 90 km. The normalization of the \texttt{bbodyrad} component is $
    N_{\rm bbodyrad}=R_{\rm km}^{2}/D_{\rm 10}^{2}$
Here, $R_{\rm km}$ is the source radius in km and $D_{\rm 10}$ is the distance to the source, and it assumes a spherical geometry of the blackbody.
Taking the normalization of $8.6_{-2.6}^{+4.0}$ for MJD 58812, and assuming that the source distance between 6.5 to 10 kpc away \citep{jana}, we can approximate a range for $R_{\rm km}$ to estimate a spherical surface area of heat propagation. Assuming that the plunging region is a thin ring at the ISCO, we collapse this spherical surface onto a ring, and find that the width of the ring $\Delta R$ may vary between 0.76 to 1.56 km. As this a physically plausible range for the ring width, we do not rule out the possibility of an extra emission component at the plunging region.
 
However, the residuals near 6 keV of the fitted spectra are still prominent. In the search for a physically sound model with a better fit, we invoke reflection of the blackbody returning radiation component. 

\subsection{Soft-state spectral results with reflection}

This analysis uses relativistic disk reflection models from the relxill model suite. We assume a
lamp-post geometry of the Comptonizing corona and use the relativistic reflection model in addition to a thermal Comptonization
component. The photon temperature from
the additional \texttt{diskbb} component is used as the seed temperature for the internal \texttt{Nthcomp} continuum. Here we consider two alternative reflection models for the outburst, one standard and the other with returning radiation. For both models, we found that the spin is unconstrained, and we therefore fixed the spin $a$ to zero. Also, as the \texttt{relxillNS} model does not interpolate below an iron abundances value of $0.5$ solar, we set that as the lower limit.

\begin{figure*}[ht]
\centering
\includegraphics[width=1\textwidth]{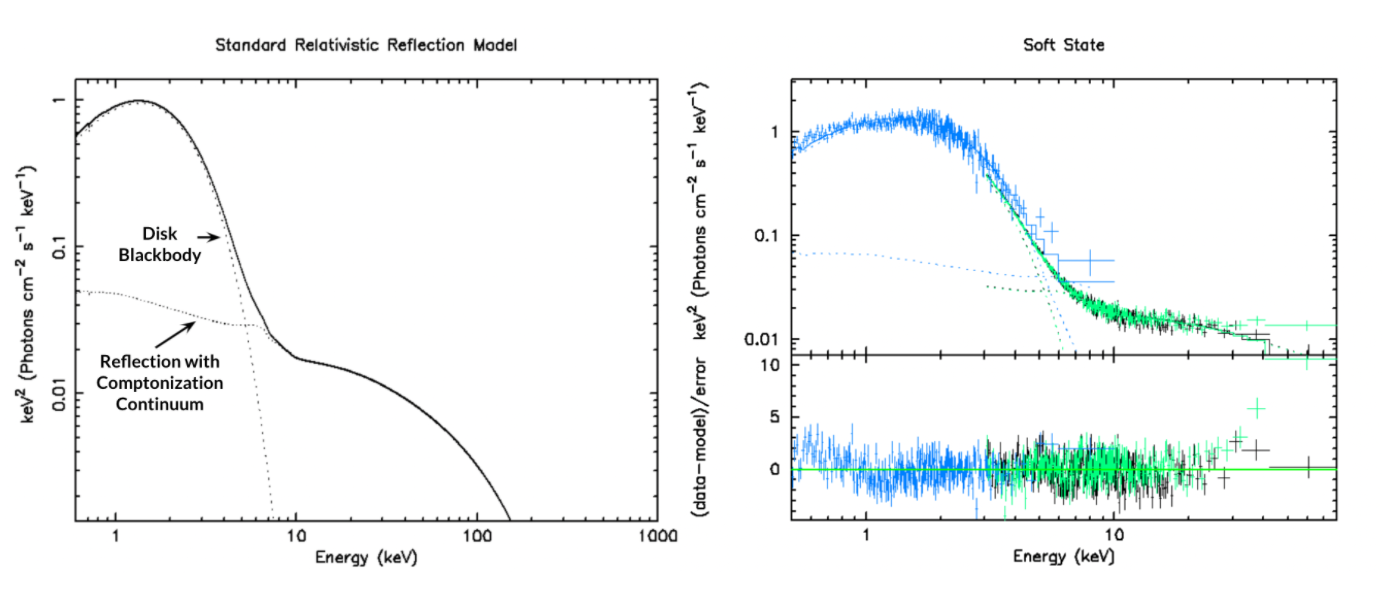}
\caption{\label{fig:s_refl} (a) The model consisting of components for a disk blackbody and reflection that includes a physical Comptonization continuum. (b) Spectra for \textit{NuSTAR} ’s 80502324006 observation and XRT’s 00012172018 observation fitted with components for a disk blackbody, thermal Comptonization, and reflection including a physical Comptonization continuum.}
\end{figure*}

At first, we added a reflection component \texttt{relxillCp}, which has an explicit \texttt{Nthcomp} component to represent the continuum, \texttt{TBabs x (diskbb + relxillCp)}. This mitigated the residuals seen for \texttt{TBabs x (diskbb + Nthcomp)} at higher energies, yielding a $\chi^{2}$ of 994 for 717 dof (see Fig. \ref{fig:s_refl}). This model combines a coronal IC spectrum atop the accretion  disk  and  the  full  ray  tracing  calculations  from the irradiating source to the disk and onward to the observer, as discussed in the introduction. The coronal IC spectrum of the \texttt{relxillCp} model is power-law-like, but the early spectra of MAXI J0637$-$430 are dominated by a softer, blackbody-like continuum.

We therefore employ the model \texttt{relxillNS}, which computes illumination of the disk by a blackbody spectrum instead of the cutoff power-law of the
original relxill model or the extra Comptonization component of \texttt{relxillCp}. The model finds values for the blackbody temperature $kT_{\rm bb}$, log of the ionization
parameter, iron abundance $A_{\rm Fe}$, log of the density of the
disk, and inclination.  
It is necessary to set a lower limit on the ionization parameter $\xi = L/nR^{2}$ where $L$ is the X-ray luminosity, $n$ is the density and $R$ is the distance from the X-ray source to the illuminated material. Taking the  distance from the source to be 6.5 to 10 kpc away, we convert the fluxes in the soft state (see Table 1) to luminosities and find that they are on the order of $10^{37}$ \,erg\,s$^{-1}$. We then take the density to be $n=1 \times 10^{19}$ cm$^{-3}$ and the distance from the source to the illuminated material to be between $10^{7}$ to $10^{8}$ cm. We obtain a lower limit of log $\xi$ $ > 3$ for the ionization parameter.  

\begin{figure*}[ht]
\centering
\includegraphics[width=1\textwidth]{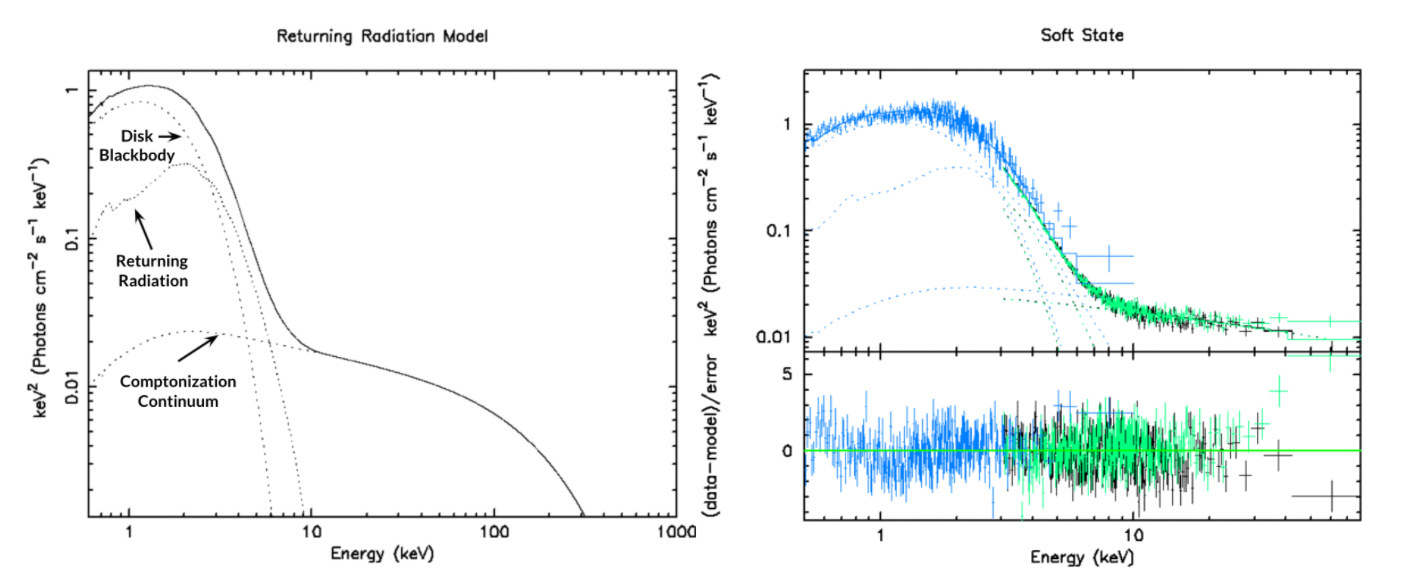}
\caption{\label{fig:006_f} (a) The model consisting of components for a disk blackbody, thermal Comptonization, and reflection of blackbody returning radiation. (b) Spectra for \textit{NuSTAR}’s 80502324006 observation and XRT’s 00012172018 observation fitted with components for a disk blackbody, thermal Comptonization, and reflection of blackbody returning radiation.}
\end{figure*}

Figure \ref{fig:006_f} displays the last model analyzed, which includes components for a disk blackbody, thermal Comptonization, and reflection of blackbody returning radiation (\texttt{TBabs x (diskbb + relxillNS + Nthcomp)}). It provides a fit with a $\chi^{2}$ of 832 for 713 dof (see Table 2). Figure \ref{fig:006_f} applies the new fit to the same soft-state observation as Figure \ref{fig:006_i}, in order to demonstrate the improvement. For a detailed breakdown of the parameters of this soft state model of our most data rich observations, see Table 3.

 We constrained the distance of the compact object by equating $R_{\rm in}$, the inner edge of the disk, with the ISCO. The normalization of the \texttt{diskbb} is given by 
     $N_{\rm disbb}=(r_{in}/D_{10})^{2} \rm cos \theta $. Here, $r_{\rm in}$ is the apparent disk radius in km, $D_{\rm 10}$ is the source distance in 10 kpc, and $\theta$ is the inclination angle. Our relationship between the inner edge of the disk and the apparent edge is given by $R_{\rm in} = \epsilon \kappa^{2} r_{\rm in}$, in which $\epsilon = 0.41$ is a factor \citep{kub} correcting the gross multi-colored disk formalism to the inner boundary condition, and $\kappa = 1.7$ - $2.0$ is
the spectral hardening factor \citep{shim}. Taking
$\kappa = 1.8$, we use the \texttt{diskbb} normalization to find that the inner radii for the observations made on MJD 58801 and 58812 are $R_{\rm in}  = (84^{+27}_{-26}) D_{10} (\mathrm{cos}\theta)^{-1/2} \: \: {\rm km}$ and $R_{\rm in}  = (77^{+24}_{-26}) D_{10} (\mathrm{cos}\theta)^{-1/2} \: \: {\rm km}$ (see Table 4).

If the compact object may be a non-rotating black hole, we can use the mean inner radius to ultimately find the mass of the BH. We therefore have

\begin{equation}
       R_{\rm in} = (81 \pm 26) D_{10} (\mathrm{cos}\theta)^{-1/2} \: \: {\rm km}
\end{equation}

\noindent For a Schwarzschild black hole, $R_{ISCO} = 6R_{\rm g}$, giving

\begin{equation}
       M = (9 \pm 3 ) D_{10} (\mathrm{cos}\theta)^{-1/2}  M_\odot
\end{equation}
\noindent 

\noindent This constraint on the $R_{\rm in}$ and $M$ is consistent with results provided by for a NICER and XRT study of MAXI J0637$-$430 outburst in the 0.5--10 keV energy range \citep{jana}. 

An advantage that the additional higher energy band \textit{NuSTAR} provides is that it allows us to detect a reflection component in the soft state, which as discussed, gives an estimation for the inclination angle. The mean inclination angle in the soft state found in this study is $64 \pm 6$ degrees, which yields

\begin{equation}
       M = (14 \pm 6) D_{10} M_\odot
\end{equation}

\noindent The distance estimate of 6.5-10\,kpc leads to a black hole mass similar to the masses found in other black hole binaries. A similar re-derived distance estimate for this analysis is found in the discussion, as well as a distance estimate for spinning black hole.  As this model gives the most optimal fits and results in plausible estimates for the inner edge of the disk and mass of the BH, we conclude that the soft state of MAXI J0637$-$430 is well-described by it. Table 2 provides a summary of the reduced $\chi^{2}$ values of the different models tested in the soft state, with a model including a \texttt{relxillNS} component being the best fit for both observations. 
\\
\subsection{Spectral results through state transition}

\begin{figure}[ht]
\centering
\includegraphics[width=0.5\textwidth]{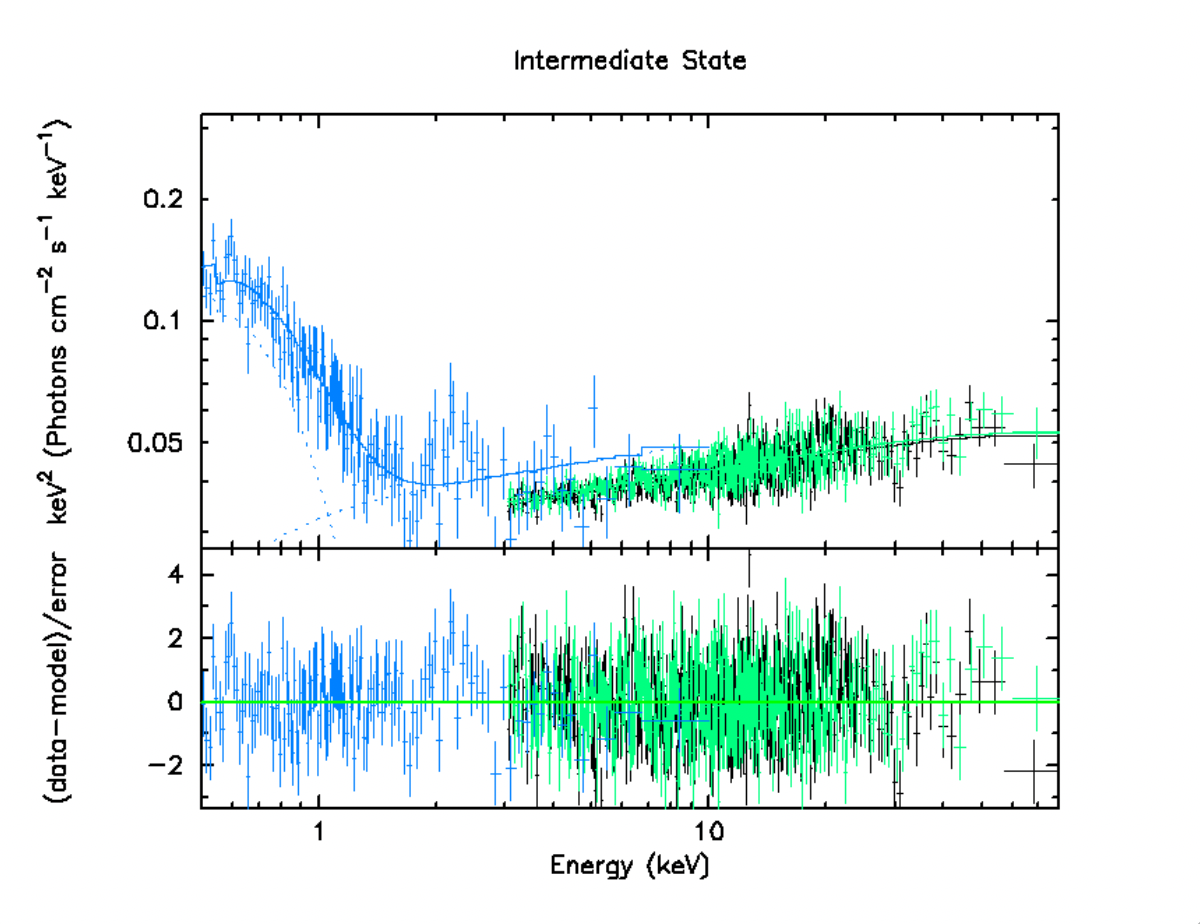}
\caption{\label{fig:008} Spectra for \textit{NuSTAR} ’s 80502324008 observation and XRT’s 00012172066 observation fitted with a disk blackbody and a thermal Comptonization component.}
\end{figure}

\begin{figure}[ht]
\centering
\includegraphics[width=0.5\textwidth]{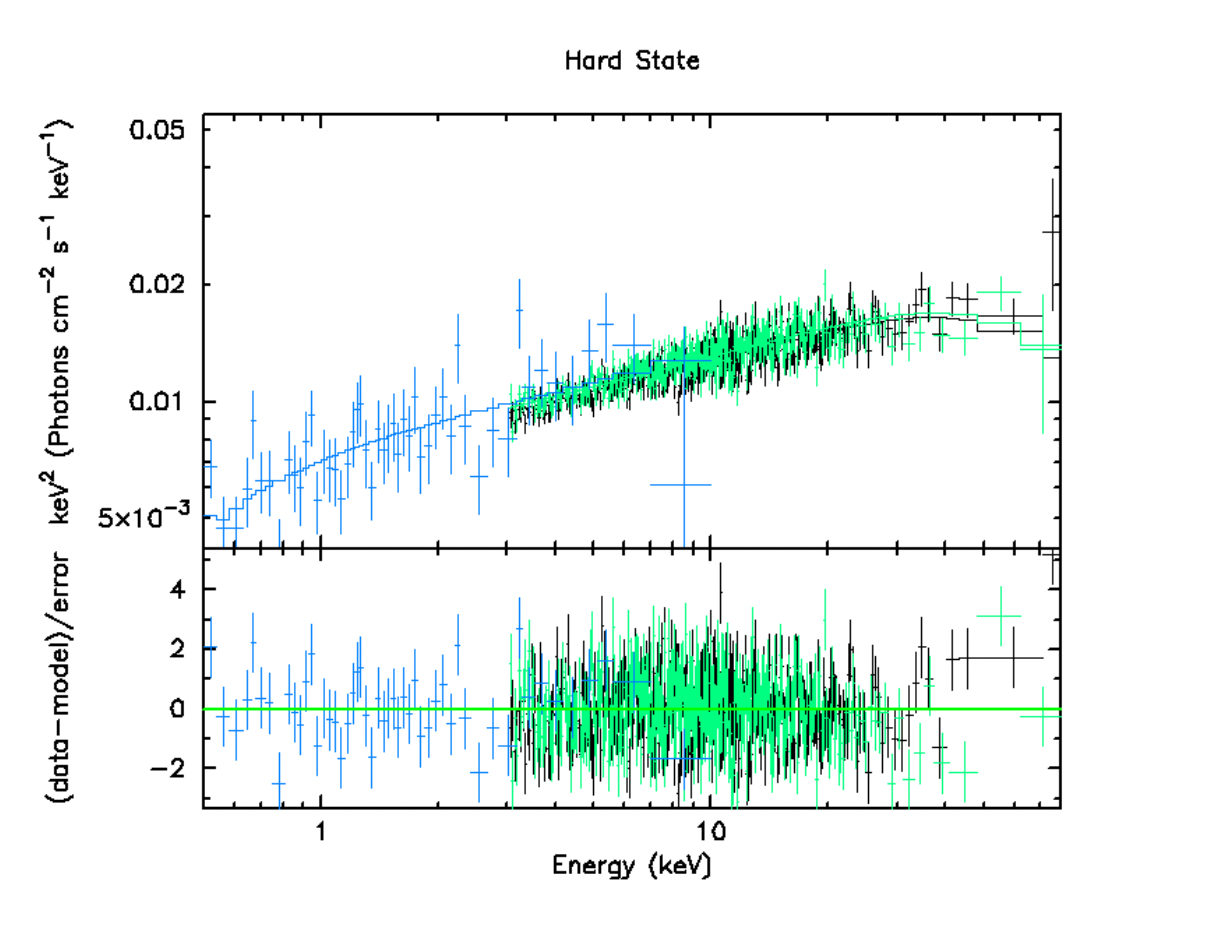}
\caption{\label{fig:010} Spectra for \textit{NuSTAR} ’s 80502324010 observation and XRT’s 00012172077 observation fitted with a thermal Comptonization component.}
\end{figure}

Throughout our observations of MAXI J0637$-$430, we  see  the  source  transition  from  a  soft  state  with  a strong  disk-blackbody component to a hard state dominated by a thermal Comptonization component. Figure \ref{fig:008} shows the outburst though state transition, in which the \texttt{diskbb} component is still quite strong, with a inner disk temperature that fell by a factor of three. Finally, in Figure \ref{fig:010}, we show that the disk component is not detectable in the hard state. 

We can use the normalization of \texttt{diskbb} once again to obtain the inner edge radius $R_{\rm in}$ of the disk at the soft-hard state on MJD 58866, which results in

\begin{equation}
       R_{\rm in} = (229^{+102}_{-116}) D_{10} (\mathrm{cos}\theta)^{-1/2} \: \: {\rm km}
\end{equation}

\noindent This shows that when blackbody reflection is no longer detected in this intermediate state, the range of the $R_{\rm in}$ increases by a factor of three while temperature drops by a factor of three (see Table 4). The inner disk temperature drop and $R_{\rm in}$ increase in the soft-to-hard transition indicate that softer reflected emission due to self-irradiation is associated with a hotter disk with a smaller inner radius. The spectral results for a BHB 4U 1630--47 outburst \citep{con} also found this association for the soft state.

\begin{figure*}
\centering
\includegraphics[width=0.8\textwidth]{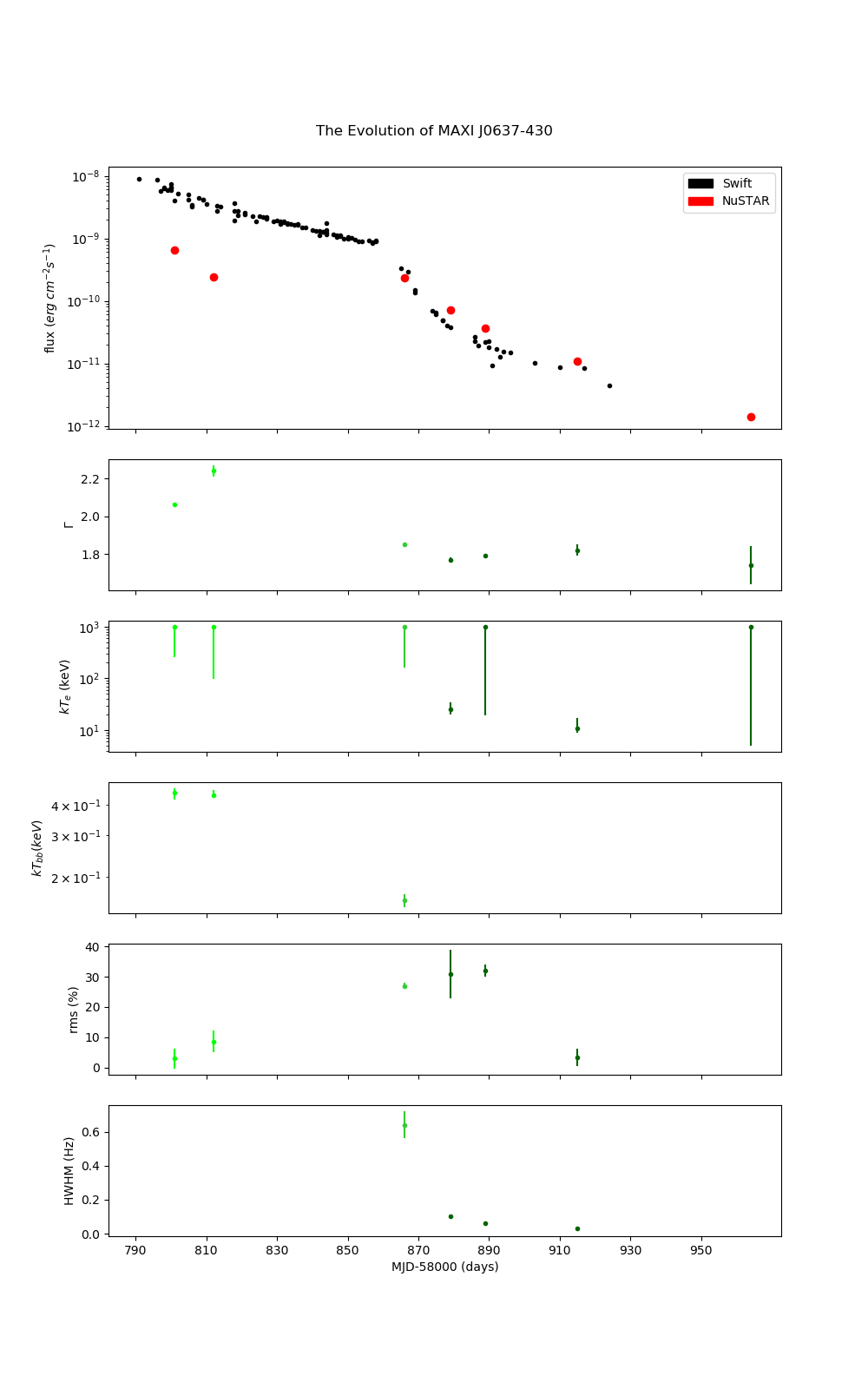}
\caption{\label{fig:evol} The evolution of (a) the source flux for XRT (0.5--10 keV) \citep{tet} and \textit{NuSTAR} (3--79 keV), (b) power-law photon index, (c) electron temperature in keV (d) the inner disk temperature in keV, (e) the rms (\%) derived from the power spectra, (f) Break frequencies from power spectra.}
\end{figure*}

The defining models used are \texttt{TBabs x (diskbb + relxillNS + Nthcomp)} for the soft state, \texttt{TBabs x (diskbb + Nthcomp)} in the transitional state, and \texttt{TBabs x Nthcomp} in the hard state. Table 1 indicates which observations correspond to each model. The state is seen to transition on MJD 58866, when the reflection component in the \textit{NuSTAR} energy range is no longer detected. The parameter values throughout the observations are summarized in Table 4 and Fig. \ref{fig:evol}.

In the hard state, we no longer detect a disk-blackbody component. We can use the Cross Power Density Spectrum (CPDS) to obtain break frequencies, which may trace the inner radius even after the thermal component falls below our bandpass. The following section details this procedure.

\section{Analyzing the Cospectra} \label{sec:timing}

Rather than studying the Power Density Spectrum (PDS) for each observation, we utilized the fact that \textit{NuSTAR} observes simultaneously with two instruments, FPMA and FPMB, in order to produce the CPDS for each observation. The CPDS is given by

\begin{equation}
    C(\nu) = \mathcal{F}^{*}_{\mathrm{A}}(\nu) \mathcal{F}_{\mathrm{B}}(\nu)
    \label{eq:CPDS}
\end{equation}

\noindent Where $\mathcal{F}^{*}_{\mathrm{A}}(\nu)$ is the complex conjugate of the Fourier transform of the light curve observed by FPMA and $\mathcal{F}_{\mathrm{B}}(\nu)$ is the Fourier transform corresponding to FPMB. The principal advantage of the CPDS as compared with the PDS is the fact that its real part, the cospectrum, only includes the power due to signals which are in phase between the two light curves, thereby eliminating contributions due to background and dead time \citep{sean}.

\begin{figure*}[ht]
\centering
\includegraphics[width=0.8\textwidth]{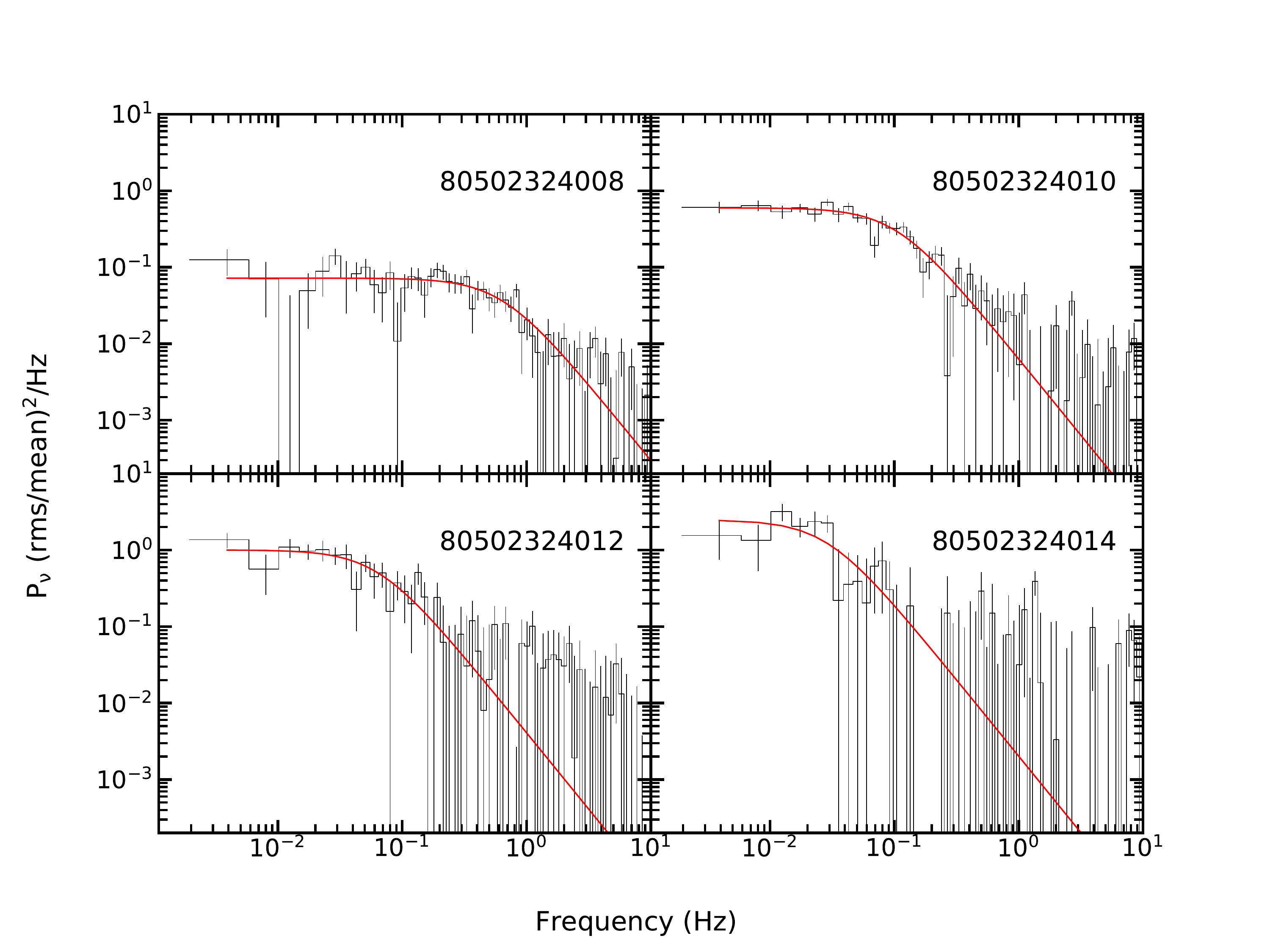}
\caption{\label{fig:cpds} The average cospectra for FPMA and FPMB for hard and soft-hard \textit{\textit{NuSTAR}} observations (MJD 58866, 58879, 58889, 58915)}
\end{figure*}

We began by shifting the arrival times in the cleaned event files to the barycenter of the solar system by calling {\tt barycorr} while extracting scientific products with \texttt{ nuproducts}. Calling {\tt barycorr} in this way defaults to the FK5 reference frame and the corresponding ephemeris DE-200. We specified the coordinates of the source in the FK5 reference frame as determined by the automatic centroid detection tool provided by DS9. In order to avoid introducing an artificial offset between FPMA and FPMB arrival times, we used the source coordinates as determined using FPMA to correct the event files for both modules. We supplied the {\em NuSTAR} clock correction file v111 generated on October 30, 2020. Using the Stingray package \citep{sean2}, we then produced FPMA and FPMB light curves with resolution 1/512 \,s, filtering events using source regions with radius $90^{\prime\prime}$ and centered on the source using the automatic centroid function provided by DS9. We did not filter for energy, instead producing light curves for the entire {\em \textit{NuSTAR}} energy range (3--79\,keV). Next, we split the light curves into segments of length $256$\,s. For each pair of segments, corresponding to FPMA and FPMB, we produced the cospectrum. All of the cospectra produced for a given observation were then averaged, resulting in the average cospectra shown in Figure \ref{fig:cpds}, which have been rebinned for clarity.

The cospectra appear relatively featureless, aside from low-frequency noise. In order to characterize the shape of this noise, we fit each cospectrum with a single zero-centered Lorentzian, described by

\begin{equation}\label{eq:Lorentz}
    P(\nu) =  \frac{A}{\pi} \left[ \frac{\gamma}{\nu^{2} + \gamma^{2}} \right]
\end{equation}

\noindent Where  $P(\nu)$ is the rms-normalized power density, $A$ is the total integrated power under the Lorentzian, and $\gamma$ is the half-width at half-maximum (HWHM). The total power, accounting only for positive frequencies, is given by $A/2$. This is related to the total rms of the component (in units of percent) by the relation $\mathrm{rms} = \sqrt{A/2}\times100$.

Figure \ref{fig:evol} and Table 4 summarize the evolution of the values derived from power spectra. We note that throughout the outburst, $R_{\rm in}$ increases and the half-width frequency decreases. Meanwhile, the power-law photon index ($\Gamma$) drops. The implications of this are further explored in the discussion.

\section{Discussion}
\subsection{Evolution of spectral and timing properties over time}

This analysis classified MAXI J0637$-$430 in three spectral states: soft, intermediate, and hard. To detail the change of properties through state transition, we provide the results of the spectral and timing analysis over time in Figure \ref{fig:evol}, showing the evolution of (a) the source flux for XRT \citep{tet} and \textit{NuSTAR}, (b) power-law photon index, (c) electron temperature of the corona in keV (d) the inner disk temperature in keV, (e) the rms (\%) derived from the power spectra, and (f) break frequencies from power spectra.

The light curve in panel (a) of Figure \ref{fig:evol} displays the flux changes in XRT's 0.5--10 keV energy band and \textit{NuSTAR}'s 3--79 keV band. \textit{NuSTAR}'s hard X-ray continuum is advantageous in identifying the presence of Comptonization, while XRT's soft X-ray continuum helps identify a disk blackbody component. Consequently, XRT observes higher fluxes in early soft-state observations compared to \textit{NuSTAR}. The shift in dominance to \textit{NuSTAR} in later observations illustrates the transition to the hard state. The \textit{NICER} campaign \citep{jana} divided the transitions into finer increments of the soft intermediate state (SIMS), high soft state (HSS), hard-intermediate state (HIMS), and low hard state (LHS) by tracking the photon power-law index and the correlation between spectral states and timing properties. We did not distinguish between the two flavors of the soft state as we did not obtain fitted power-law photon indices indicating a transition. As for later states, our intermediate state observation occurred when \citet{jana} indicate that the source was in the HIMS.

A combination of the spectral analysis in the soft and intermediate state and timing analysis in the hard state makes it possible to map out the evolution of the accretion disk geometry throughout the outburst.  
 The disk-blackbody component in the soft and intermediate state (sections 3.2 and 3.3) indicate that the range of the inner disk radius increases by a factor of three. Once the source is in the hard state and the disk-blackbody component is no longer detected, we use the properties of the power spectrum to constrain the inner radius following \citet{dimatteo}. 
 We therefore then use the Lorentzian half-width frequency to trace the inner disk radius once the disk-blackbody component is no longer detectable in the spectra. The Lorentzian half-width frequency is found to decrease from the intermediate to hard state by a factor of 21. We can place an upper bound on the increase of the inner radius from Keplerian orbits using $R \propto \nu^{-2/3}$, where $\nu$ is the characteristic frequency. Therefore, the half-width frequency decrease implies an increase of inner radius by a factor of 8 from the intermediate to hard state. 
 
 The relationship between coronal temperature, power-law index, and inner radius (summary in Table 4) is best understood by models consisting of an inner optically thin corona and an outer optically thick radius. In the early stages of the outburst, we see that the power-law index decreases as the corona temperature decreases. This may be because soft photons of a disk cool the corona, producing a softer power-law index. However, as the mass accretion rate drops, the inner disk radius moves away from the compact object, and the corona is subject to a lower flux of soft photons from the disk. This lower influx of soft photons causes the coronal spectrum to harden.
The correlation between a decreasing half-width frequency and increasing power-law photon index, previously identified in the hard states of other black hole systems \citep{gil, rev, kal, freq}, is also seen in this outburst for MAXI J0637$-$430.

\subsection{Nature of the compact object of MAXI J0637$-$430}



We can make further inferences on MAXI J0637$-$430 by deriving the compact object mass and distance from the inner disk radius.  For a non-rotating compact object, we derive a mass of  $M = (9 \pm 3 ) D_{10} (\mathrm{cos}\theta)^{-1/2}  M_\odot$ from the assumption that the ISCO lies at 6 $R_{\rm g}$ for the non-rotating Schwarzschild black hole. However, as the spin is unconstrained in our spectral analysis, we also consider the possibility that the MAXI J0637$-$430 compact object may be rotating and that the effects of spin are not evident in our spectra. In the extremal Kerr case, theoretical support and decades of empirical evidence motivate linking $R_{ISCO}$ to $R_{in}$ \citep{ste}. The mass could therefore be up to six times larger than for the Schwarzschild assumption.  Although we discuss the source distance further below, the lower limit of 6.5\,kpc derived by \cite{jana} corresponds to $M > 6.0\pm 2.0 M_\odot$, which would require the presence of a black hole.  The least massive black hole found so far has M = 3.3 $M_\odot$ \citep{low}, while neutron stars reach an upper range between 1.5 to 3  $M_\odot$.

Although the extremal Kerr case still qualifies MAXI J0637$-$430 as a stellar black hole when we consider mass alone as an indicator, it is necessary to consider the implication it would have on distance. For a non-rotating black hole, we estimate distance by solving a system of two equations between the mass-distance relation derived in section 3.2 and a fraction of the state-transition luminosity \citep{jana}. To obtain the state-transition luminosity, we extrapolated the flux between XRT observations 00012172060 and 00012172064, obtaining $1.15 \times 10^{-9}$ erg cm$^{-2}$s$^{-1}$ for 0.5--10 keV, and multiplied it by $4 \pi D_{\rm 10}^2$. Through a thorough study, \citet{mac} demonstrated that the state-transition luminosity for a LMXB typically falls between 0.01 and 0.04 of the Eddington limit, which is given by $L_e = 1.3 \times 10^{38} (M/M_\odot)$\,erg\,s$^{-1}$. Solving 
\begin{equation}
\begin{split}
    1.15 \times 10^{-9} \: \: {\rm erg \: cm}^{-2}{\rm s}^{-1} \times 4 \pi D_{\rm 10}^2 = \\ 0.01 \times 1.3 \times 10^{38} (M/M_\odot)\, \: \: {\rm erg} \: {\rm s}^{-1}
\end{split}
\end{equation}
\noindent with the mass-distance equation (2) derived from \texttt{diskbb} gives a distance between 5.9 and 11.5 kpc for an inclination of $\theta = 0$.

In the extremal Kerr scenario, our mass-distance equation is $M_{\rm in} = (54 \pm 18 ) D_{10} (\mathrm{cos}\theta)^{-1/2}  M_\odot$. This would imply a distance of at minimum 33 kpc. We consider that 
MAXI J0637$-$430 is at a galactic latitude --20 degrees. This would place the compact object at least 11 kpc from the galactic disk. As the galactic disk is the 
birthplace of supergiant and massive stars that evolve into neutron stars or black holes, this would be unusual indeed. If the disk inclination is at the measured average of $\theta = 64$ degrees, or if the state-transition luminosity is higher than 0.01 of the Eddington limit, it could qualify as a galactic halo object. A non-rotating or low-spin case is more likely than this rare scenario. 

It would be remiss not to consider the possibility of a neutron star as a compact object. Looking at the equation of $R_{\rm in} = (81 \pm 26) D_{10} (\mathrm{cos}\theta)^{-1/2} \: \: {\rm km}$ in the soft state, taking a fiducial neutron star radius of 10 km, and liberally setting $\theta = 0$, the distance would be 2.0 kpc. This close range seems highly improbable considering the extremely low luminosity detected in the soft state, implying a 0.9\% Eddington luminosity for a $1.4 M_\odot$ NS. This unlikely scenario seems to favor the BHXRB interpretation. 

We draw comparisons to a study by \citet{cac} on reflection from the inner accretion disk around neutron stars. The disk blackbody norm for nine neutron stars was found to be on the order of 100 to 200, with a high inner disk temperature on the order of 1 or 2 keV. MAXI J0637$-$430 does not mimic this consistent pattern of a notably hotter disk with a smaller inner radius, and instead exhibits a cool disk temperature as described by \citet{rem}. The magnitudes of the inner disk normalization and temperature are comparable to the results of a returning radiation fit for BHXRB XTE J1550–564 \citep{ex}. 

\subsection{Broadband spectrum and concluding remarks}

While previous studies found that the soft states of the MAXI J0637$-$430 outburst are well-described by a two-component model, \textit{NuSTAR}'s broader bandpass at 3--79 keV improves the quality of the spectrum and introduces positive residuals in the iron line region and above 10 keV. We tested both excess emission from the plunging region and two reflection scenarios. We conclude that excess emission from the plunging region yields a physically plausible width of a thin ring at the ISCO. However, we found that including a returning radiation reflection component gives the best fits, while also providing physically plausible estimates for the inner edge of the disk and mass of the black hole. 

In testing reflection components, we found that the spectra of MAXI J0637$-$430 favor the softer,  blackbody-like spectrum of the \texttt{relxillNS} model over the power-law-like coronal IC spectrum of the \texttt{relxillCp} model. The addition of this feature causes some changes worth noting. As an extra emission component is accounted for in the \texttt{relxillNS} model, it drops the temperature of the inner disk blackbody when compared to models that omit reflection (as seen by comparing to the 0.5--10 keV analysis by \citep{tet}. Although the residuals near the iron line were originally interpreted as iron emission features \citep{fe}, they were not detected in this analysis.  This may be due to the low iron abundance ($A_{\rm Fe}$) value derived from our spectral fits, which give a best fit at the minimum value used in creating the \texttt{relxillNS} model (0.5 solar). This analysis demonstrates that a reflection component describes the spectra well despite the iron line non-detection.

It is interesting to note that typically halo stars originating from globular clusters have lower metallicities than stars in the Galactic plane, so a black hole accreting material from a halo companion star would be expected to have a low iron abundance. Regardless of this indicator, MAXI J0637$-$430 as a galactic halo object would still be unusual. If MAXI J0637$-$430 has a black hole as a compact object, a massive star progenitor would be required, which is rare in the halo.

To gain a further understanding of accretion efficiency, or the fraction of energy actually radiated of that available according to the mass transfer rate into the accretion disk, we can calculate the ratio of our observed luminosity to the Eddington limit. Taking the high source flux on MJD 58801 for 0.5--10 keV, assuming a source distance of 8.7 kpc, we find $L_{obs}/L_{e} = 0.029$, a relatively low value for a soft state \citep{mac, mot}. The ionization parameter found by \texttt{relxillNS}, which is low in the context of a BHXRB, can explain this low fraction of observed luminosity.

Though the reflection spectrum of \texttt{relxillNS} gives the best fits, we emphasize that in the soft state, we find that the ratio of the flux of the reflection component and the flux of the disk blackbody component is an average of $\sim$ 0.6. \citet{cunning} demonstrates that a returning radiation component is prominent for objects of spins between 0.9-0.998, and thus this flux ratio may cast doubt on the non-rotating scenario. But as explained in the section above, a high spin would imply an unusually large distance from the galactic disk. This disconnect presents a challenge to the returning radiation scenario in this case and motivates further studies on returning radiation flux prominence for spins below 0.9.

We have proposed two possible models for the soft-state X-ray spectrum of MAXI J0637$-$430, i.e., emission from the plunging region or reprocessing of returning disk radiation. Although the present study cannot confirm or rule out any of these two models, perhaps the most important conclusion is that the broadband spectra in the soft state are not adequately described by a two-component model based on the ``standard” power-law continuum and associated reflection spectrum, such as in previous studies. The complexities observed in our data are not evident in a more limited energy bandwidth. This situation is not unique to MAXI J0637$-$430, as it has been reported in studies of other sources in their soft state (e.g., MAXI J1820+070, \citealt{plunge}; XTE J1550$-$564, \citealt{ex}; EXO 1846$-$031, \citealt{wang}; BHB 4U 1630--47 outburst, \citealt{con}). 

Moreover, the validity of one model does not necessarily rule out the other, as it is possible that the most suitable model is a combination of both the plunging region and returning radiation scenarios. The plunging region resides between the ISCO and event horizon, and thus its emission diminishes as the ISCO nears the event horizon, such is the case as maximal spin is approached. It is also feasible to postulate that some photons from the inner disk are bent back by the strong gravity of the black hole and reflected off the disk surface, but the fraction of photons experiencing this phenomenon depends strongly on the spin and geometry of illumination. A high fraction of returning photons is expected to occur at high spin. As the measurements throughout the outburst gave unconstrained values for spin, we derived the extremal cases of the mass-distance equation based on assumptions of the location of the ISCO, which depends on the extremal spin cases. Setting limits on distance with quiescent measurements could in turn better constrain the spin of the system. We therefore recommend quiescent optical and near-IR coverage data to be collected in order to better inform the spin of the system, which would help in determining the contribution of different model components. This will also help in evaluating the rare possibility of a distant halo object in addition to developing a model. Nonetheless, the options presented here are valuable when putting together a picture of MAXI J0637$-$430, as the broadband spectra require complex models that invoke physics of the inner accretion disk close to the complex object.

\acknowledgements

This work made use of data from the \textit{NuSTAR} mission, a project led by the
California Institute of Technology, managed by the Jet Propulsion Laboratory, and funded by the National Aeronautics and Space Administration. We thank the \textit{NuSTAR} Operations, Software and  Calibration teams for support with the execution and analysis of these observations.  This research has made use of the \textit{NuSTAR}  Data Analysis Software (NuSTARDAS) jointly developed by the ASI Science Data Center (ASDC, Italy) and the California Institute of Technology
(USA).  This work also made use of data from the \textit{Neil Gehrels Swift Observatory}.  H.L. and J.A.T. acknowledge partial support under NASA \textit{NuSTAR} Guest Investigator grant 80NSSC20K0644. J.H. acknowledges support from an appointment to the NASA Postdoctoral Program at the Goddard Space Flight Center, administered by the Universities Space Research Association under contract with NASA. J.A.G. acknowledges support from NASA NuSTAR Guest Investigator grant 80NSSC20K1238, and from the Alexander von Humboldt Foundation. J.J. acknowledges support from the fellowship of China Postdoctoral Science Foundation (2021M691822), the Tsinghua Shuimu Scholar Program and the Tsinghua Astrophysics Outstanding Fellowship. Thank you to Dr. Thomas Dauser and Dr. Jack Steiner for the useful comments along the way.

\smallskip
\facilities{\textit{NuSTAR}, \textit{Swift}/XRT}

\smallskip
\software{XSPEC (v12.11.1c \citealt{xspec}), HEAsoft (v6.27.2), relxill suite} \citep{daus, Garc}
  
\clearpage


\appendix
Every model tested produces positive residuals in the $<$ 1 keV range (see Fig. \ref{fig:plunge}, \ref{fig:s_refl}, and \ref{fig:006_f}). As the \cite{tet} optical spectroscopy analysis carefully derived the hydrogen column density from Ultraviolet and Optical Telescope (UVOT) measurements of MAXI J0637$-$430, we fixed it to the hydrogen column density of $4.39 \times 10^{20}$ cm$^{-2}$. Similarly, HI4PI find a Galactic column density in the direction of the source $5.23 \times 10^{20}$ cm$^{-2}$. However, without knowing the distance to the source, it could be argued that the global dust or neutral atomic hydrogen column density estimates determined by HI4PI can only serve as an upper bound. 

One option for the lower-energy residuals could be variation in elemental abundances in the interstellar medium. To test this, we fit to the model \texttt{tbnew\_feo x (diskbb + relxillNS + Nthcomp)}, in which \texttt{tbnew\_feo} from the absmodel \footnote{https://pulsar.sternwarte.uni-erlangen.de/wilms/research/tbabs/} package is a form of the \texttt{Tbabs} component that considers variation in oxygen and iron abundances. Taking the column density to be $4.39 \times 10^{20}$ cm$^{-2}$, and tying the oxygen and iron abundances together, we find that the reduced  $\chi^{2}$ shows only slight improvement for MJD 58801 at 825 for 714 dof, with the <1 keV residuals slightly smaller still present. The two elemental abundances fit to a subsolar value of $<0.6$  solar.

\begin{figure}[ht]
\centering
\includegraphics[width=0.52\textwidth]{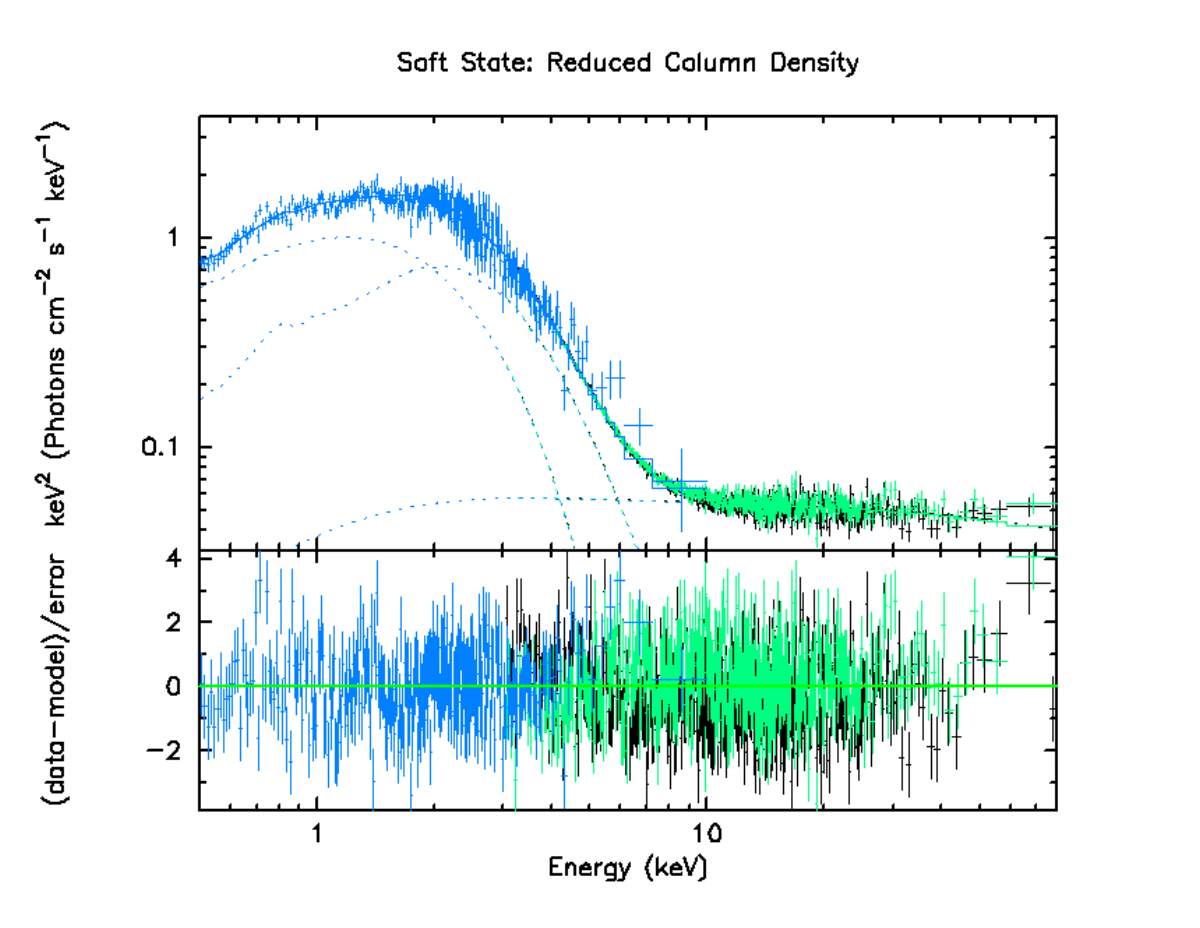}
\caption{\label{fig:nH} Spectra for \textit{NuSTAR} ’s 80502324004 observation and XRT’s 00012172008 observation fitted with components for a disk blackbody, thermal Comptonization, and reflection of blackbody returning radiation. Column Density converged to $2.26 \times 10^{20}$ cm$^{-2}$ }
\end{figure}

To check whether a lower value of the column density eliminates the low-energy residuals, we repeated the analysis for the soft state and found the column density from the fit to the XRT and \textit{NuSTAR} spectra. For the most statistically robust observation (MJD 58801), the returning radiation model gave that the column density is equal to $2 \pm 1 \times 10^{20}$ cm$^{-2}$, and the plunging region model gave that the column density is $3_{-2}^{+4} \times 10^{20}$ cm$^{-2}$. The $\chi^{2}$ value remained relatively the same for the plunging region model at 2158 for 1156 dof, and improved slightly for the returning radiation model at $\chi^{2}$ of 1311 for 1151 dof. The residuals below $<$ 1 keV were diminished (Fig. \ref{fig:nH}). We note that the change of column density did not alter the parameter values outside the error ranges listed in Table 3 and for the $R_{\rm in}$ values in Table 4.  

The change of column density slightly reduces the ratio of the flux of the reflection component and the flux of the disk blackbody component, summarized in Table 5. For either a column density fixed to $4.39 \times 10^{20}$ cm$^{-2}$ or found from the fitted spectra, the ratio is notably high for the earlier observation on MJD 58801. If the column density is $4.39 \times 10^{20}$ cm$^{-2}$ or higher, than there may be an extra emission component in the early stages of the soft state currently unaccounted for in all models.



\begin{table}
\centering
\caption{Observations\label{tab:obs}}
\begin{minipage}{\linewidth}
\centering
\footnotesize
\begin{tabular}{cccccccc} \hline \hline
MJD &  \textit{NuSTAR} & \textit{NuSTAR} Flux (3--79 keV) &\textit{NuSTAR} Exposure& XRT  &  XRT Flux (0.5--10 keV) &XRT Exposure& Model
\footnote{(1) soft state: \texttt{TBabs x (diskbb + relxillNS + Nthcomp)} \\ (2) transition: \texttt{TBabs x (diskbb + Nthcomp)} \\ (3) hard state: \texttt{TBabs x Nthcomp}} 

\\ & ObsID  & $ \times 10^{-10}$ erg cm$^{-2}$s$^{-1}$ & Time (s) & ObsID&$\times 10^{-10}$ erg cm$^{-2}$s$^{-1}$ & Time (s) & Fit
\\ \hline \hline
58792  & 80502324002 &8.19 $\pm$ 0.02& 36799 & --- & --- & --- &1    \\
58801  & 80502324004 &6.44 $\pm$ 0.02& 67738 & 00012172008 & 42.23 $\pm$ 0.17  & 2525 & 1 \\
58812 & 80502324006 &2.39 $\pm$ 0.02& 48626 & 00012172018 & 30.75 $\pm$ 0.78 & 1667 & 1 \\
58866  & 80502324008 &2.37 $\pm$ 0.02& 46630 & 00012172066 & $ 2.98 \pm 0.07 $ &674& 2 \\
58879 & 80502324010 &0.72 $\pm$ 0.01 & 110779  & 00012172077& $0.46 \pm 0.02$ &1686& 3 \\
58889 & 80502324012 &0.37 $\pm$ 0.01& 50233 & 00012172085 &  $0.17 \pm 0.01$ & 1860&3           \\
58915  & 80502324014 &0.11  $\pm$ 0.01& 65379 & 00012172093 & 0.06 $\pm$ 0.02  & 944&3 \\ 
58964  & 80502324016 &0.014 $\pm$ 0.001& 47486 & --- & --- & --- &3 \\
\hline
\end{tabular}
\end{minipage}
\end{table}

\begin{table}
\caption{Testing Soft State Fits\label{tab:comp}}
\begin{minipage}{\linewidth}
\centering
\footnotesize
\begin{tabular}{ccc} \hline \hline
Model &   $\chi^{2}$/dof & $\chi^{2}$/dof \\
& MJD 58801 & MJD 58812 \\\hline \hline
 \texttt{TBabs x(diskbb + Nthcomp)}& 2374 / 1158 & 1242 / 720 \\
 
 \texttt{TBabs x(diskbb + Nthcomp + bbodyrad)}&   2154 / 1156  & 859 / 718  \\

  \texttt{TBabs x(diskbb + relxillCp)}  & 1901 / 1155 & 944 / 717 \\
 
 \texttt{TBabs x(diskbb + relxillNS + Nthcomp)}   & 1320 / 1151 & 832 / 713 \\

\hline
\end{tabular}
\end{minipage}
\end{table}

\begin{table}
\caption{Soft State Model\label{tab:soft} \footnote{Errors are reported at the 90\% confidence level and calculated. There were no distinctions made for the emissivities for the coronal flavor models in the inner and outer disk, so they were fixed such that $Index_{1}=Index_{2}=3$. The
dimensionless spin parameter and redshift were set to zero for the
RELXILLNS model.}}
\begin{minipage}{\linewidth}
\centering
\footnotesize
\begin{tabular}{cccc} \hline \hline
Model &  Parameter &  MJD 58801 & MJD 58812  \\ \hline \hline
 \texttt{TBABS}& $N_{\rm H} (10^{22}$cm$^{-2})$ & 4.39 $\times 10^{-2}$ &  4.39 $\times 10^{-2\dagger}$ \footnote{$\dagger$ is fixed for all observations}\\
 
\texttt{DISKBB} &   $kT_{\rm in}$ (keV)  & $ 0.45_{-0.03}^{+0.02}$ & $0.44_{-0.01}^{+0.02}$  \\

  & norm  & $3964_{-378}^{+410} $& $3312_{-384}^{+319}$\\
 
\texttt{RELXILLNS}   & $i^{(\circ)}$ \footnote{a hard limit of 70 degrees was placed} &$ 70_{-1}$ &$ 58_{-7}^{+7}$  \\

  & $kT_{\rm bb}$  (keV)&  $0.523_{-0.002}^{+0.003}$ & $0.529_{-0.003}^{+0.010}$ \\
 
  & $log \xi$ \footnote{a hard limit of 3 was placed, see section 3.2}  &$ 3^{+0.01}$ & $ 3^{+0.01}$  \\
 
 &  $A_{\rm Fe}$ \footnote{a hard limit of 0.5 was placed}  & $ 0.5^{+0.04}$ & $0.5^{+0.06} $\\

 &  $log N$  & $17.3^{+0.4}_{-0.3}$ & $17.3^{+0.6}_{-0.2} $ \\
 
  &  $norm(10^{-5})$  & $21^{+8}_{-0.3}$ & $8^{+4}_{-1} $ \\
        
\texttt{NTHCOMP}  &$ \Gamma $& $2.06_{-0.01}^{+0.01} $& $2.24_{-0.03}^{+0.03}$ \\
   & $kT_{\rm e}$ (keV)   & $>$ 257 &  $>$ 98 \\
      & $kT_{\rm in}$ (keV) \footnote{This value is tied with the $T_{\rm in}$ value in \texttt{DISKBB}} & tied & tied \\
   & norm   & $0.042_{-0.002}^{+0.002} $&  $0.019_{-0.002}^{+0.001}$ \\
   
\hline
\end{tabular}
\end{minipage}
\end{table}

\begin{table}
\caption{Fit Parameter Summary\label{tab:summary}}
\begin{minipage}{\linewidth}
\centering
\footnotesize
\begin{tabular}{ccccccccc} \hline \hline
MJD & state & \texttt{DISKBB} norm & $kT_{\rm in}$ (keV) & $kT_{\rm e} $ (keV) & $\Gamma$ & $R_{\rm in}$ (km) \footnote{$D_{\rm 10}$ is the source distance in 10 kpc, $\theta$ is inclination angle.} & FWHM/2 \footnote{The half-width of the zero-centered Lorentzian in Hz.} & rms (\%) \\ \hline \hline

 58792& soft & --- &  ---  & $>$ 114    & $2.09_{-0.01}^{+0.01}$ & --- & --- & --- \\
 
 58801& soft& $3964_{-378}^{+410}$ &  $0.45_{-0.03}^{+0.02}$  & $>$ 257    & $2.06_{-0.01}^{+0.01}$ & $(84^{+27}_{-26}) D_{10} (\mathrm{cos}\theta)^{-1/2}$ & --- & ---\\

 58812 & soft& $3312_{-384}^{+319}$ & $0.44_{-0.01}^{+0.02}$ & $>$ 98  & $ 2.24_{-0.03}^{+0.03}$ & $(77^{+24}_{-26}) D_{10} (\mathrm{cos}\theta)^{-1/2}$ & --- & ---\\
 
 58866 & intermediate & $29689_{-5922}^{+7621}$  & $0.16_{-0.01}^{+0.01}$ & $>$ 160  &  $1.85_{-0.01}^{+0.01} $ & $(229^{+102}_{-116}) D_{10} (\mathrm{cos}\theta)^{-1/2} $ & 0.64 $\pm$ 0.08 & 27 $\pm$ 1 \\

 58879 &hard& --- &     --- & $25_{-5}^{+10} $ & $1.77_{-0.01}^{+0.01}$ & --- & 0.10 $\pm$ 0.01 & 31 $\pm$ 8\\
 
 58889  & hard&--- & --- &  $>$ 19 &$1.79_{-0.01}^{+0.01}$&--- & 0.06$\pm$ 0.01 & 32 $\pm$ 2 \\
 
 58915 &hard& --- &  --- & $11_{-2}^{+6}$ & $1.82_{-0.03}^{+0.03}$ &--- & 0.03  $\pm$ 0.01 & 33 $\pm$ 3 \\

 58964 &hard& --- & ---    & $>$ 5 & $1.74_{-0.10}^{+0.10}$ &--- & --- & ---\\

\hline
\end{tabular}
\end{minipage}
\end{table}

\begin{table}
\caption{Reflection/Disk-Blackbody Flux Ratio in the Soft State \label{tab:comp}}

\centering
\footnotesize
\begin{tabular}{ccc} \hline \hline

N$_{\rm H}$ & MJD 58801 & MJD 58812 \\\hline \hline
 
 $4.39 \times 10^{20}$ cm$^{-2}$ &   0.8  & 0.4  \\

 $1.47-2.62 \times 10^{20}$ cm$^{-2}$  & 0.7 & 0.3 \\
\hline
\end{tabular}

\end{table}


\clearpage

\end{document}